\documentstyle[aps,prl,epsf]{revtex}
\draft
\begin{document}
\input{epsf}
\twocolumn[\hsize\textwidth\columnwidth\hsize
\csname@twocolumnfalse\endcsname
\title{Crystallographic structure of ultrathin Fe films on Cu(100)}
\author{Albert Biedermann, Rupert Tschelie$\ss{}$nig, Michael Schmid, and Peter Varga}
\address{Institut f\"ur Allgemeine Physik, Vienna University of Technology,
A-1040 Vienna, Austria} \maketitle

\begin{abstract}
We report bcc-like crystal structures in 2--4 ML Fe films
grown on fcc Cu(100) using scanning tunneling microscopy.
The local bcc structure provides a straightforward
explanation for their frequently reported outstanding
magnetic properties, i.e., ferromagnetic ordering in all
layers with a Curie temperature above 300 K. The
non-pseudomorphic structure, which becomes pseudomorphic
above 4 ML film thickness is unexpected in terms of
conventional rules of thin film growth and stresses the
importance of finite thickness effects in ferromagnetic
ultrathin films.

\end{abstract}
\vskip2pc]

Both academic interest in novel nanomagnetic phenomena as
well as their technological importance for
magneto-electronics and high density magnetic storage
devices make the study of ultrathin ferromagnetic films
particularly worthwhile. These extremely thin films,
typically less than 10 monolayers (ML) thick, exhibit
significantly different magnetic properties in contrast to
the bulk material, e.g., different magnetization
directions, enhanced magnetic moments, and lower Curie
temperatures. Fe films epitaxially grown on Cu(100) are
distinguished by a particular complex behavior since they
are variable both with respect to magnetic ordering (ferro-
or antiferromagnetic) and crystal structure (fcc or bcc).
Although the epitaxial system Fe/Cu(100) is under intense
scrutiny for more than a decade and its magnetic properties
have been mapped very precisely, no conclusive overall
picture of the relation between structure and magnetic
states has emerged yet. Regarding the model for films
deposited at room temperature presently discussed in the
literature, there is clear evidence for an
antiferromagnetic, pseudomorphic fcc phase between 5 and 10
ML film thickness. The character and origin of the
ferromagnetic phase between 2 and 4 ML, however, remains
unclear. Low energy electron diffraction (LEED)
\cite{mull95}, surface extended X-ray absorption
fine-structure (SEXAFS) \cite{magn91}, and medium energy
ion scattering studies (MEIS) \cite{bart95} indicate a
distinct distortion of the fcc lattice in 2--4 ML films.
This reconstruction, which is considered to comprise the
entire film thickness \cite{mull95}, is accompanied by a
substantial increase of the film volume (interlayer
distance) by about 5\% \cite{mull95,clar87}. In the past,
these results led to the notion of a second, ferromagnetic
fcc-like phase with an expanded film volume, i.e., a face
centered tetragonal (fct) phase. While {\em ab-initio}
calculations of bulk Fe do support the possibility of a
ferromagnetic fcc phase with expanded volume under
substantial tensile strain \cite{moru86}, the respective
calculations of ultrathin films on Cu(100) did not provide
an unambiguous confirmation of the ferromagnetic fct model
\cite{moro99,spis00}. This is an important question since
the hypothetical existence of a ferromagnetic fcc-like
phase is relevant also for the solid state physics of Fe.
The two--$\gamma$--state model introduced by Kaufman,
Clougherty, and Weiss \cite{kauf63}, assuming two fcc
states of bulk Fe either ferro- or antiferromagnetic, is
still under discussion.

In this Letter, we resolve this issue by characterizing the
atomic structure of the ultrathin films below 5 ML
thickness locally by scanning tunneling microscopy (STM).
The key result is the characteristic bcc-like nature of the
films, which pinpoints the true source of ferromagnetism in
ultrathin Fe films on Cu(100), and reveals the actual root
of this phenomenon: The remarkable stability of a
non-pseudomorphic bcc-like phase in a film only 2--4 ML
thick, while above 4 ML the pseudomorphic fcc structure is
more stable.

The Fe films were grown on a (100) oriented Cu single crystal at
310 K (except the 2.5 ML film grown at 130 K and annealed to 300
K) by evaporation from tips of Fe wires in ultra-high vacuum
using a deposition rate around 1 ML/min. The evaporator was
calibrated using a quartz crystal microbalance and the film
thickness checked by quantitative Auger electron spectroscopy
\cite{cump97}. Imaging was done using either a room temperature
scanning tunneling microscope (STM) or a low temperature STM
operated at 80 K with {\em in situ} sputtered W tips. Ion
scattering experiments show that the surfaces of films more than
3 ML thick deposited at room temperature contain less than a few
percent copper \cite{detz93}.

Before we report on the atomic structure of the films, a general remark: It is
well known from previous experiments that the atomic structure can change
within fractions of a monolayer and also with temperature \cite{mull95,zhar97}.
Since it is very difficult to obtain atomic resolution on these surfaces, we
did not attempt to outline the boundaries in the phase diagram by STM, which
has been done before by means of LEED and the surface magneto-optical Kerr
effect \cite{mull95,thom92,li94}, but rather focused on the principal
structures and driving forces, which support ferromagnetism in the 2--4 ML
films.

After preparation of several films in the thickness range between 2 and 4 ML,
and resolving the surface atomically either at 300 or 80 K we always observed a
rather complex microdomain pattern of $(1\times{}4)$, $(1\times{}5)$, and
$(1\times{}6)$ structures coexisting with remnants of the fcc structure. The
STM images in Fig.~1 reveal the atomic structure of the respective unit cells.
All structures can be imagined as resulting from shearing the fcc lattice by
14$^\circ$, which makes the local atomic arrangement very similar to that

\begin{figure}
\centerline{\epsfxsize=0.85\hsize\epsffile[30 120 580 750]{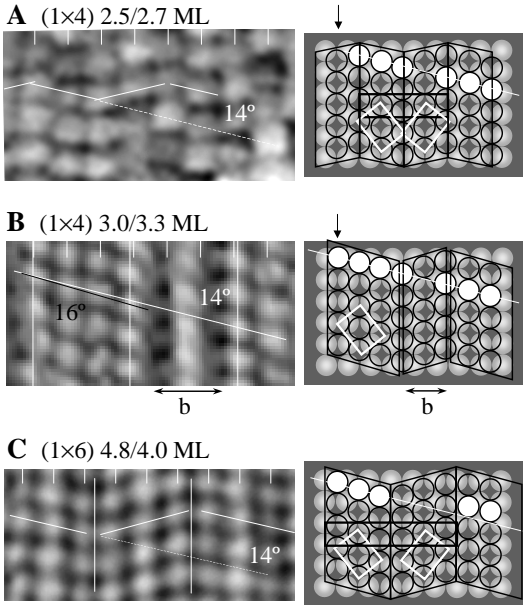}}
\caption{Atomically resolved STM images of 2--4 ML films. (A) $(1\times{}4)$
structure; Film grown at 80 K, annealed at 300 K, and imaged at 80 K. (B)
$(1\times{}4)$-like structure with boundary ``b''; Film grown at 310 K and
imaged at 80 K. (C) $(1\times{}6)$ structure; Film grown at 310 K and measured
at 310 K. The width of the either $+14^\circ{}$ or $-14^\circ{}$ sheared
bcc-like stripes varies between 2 and 4 atom rows. The first thickness value
given was measured with a quartz crystal microbalance, the second by AES.}
\label{zigzagstm}
\end{figure}

\setlength{\parindent}{0pt} of a (110) oriented bcc film in the Pitsch
orientation with respect to the fcc substrate \cite{Pits59} (see white unit
cells in Fig.~1). The atomic surface density, however, is equal to that of the
fcc(100) substrate, indicating a substantial strain in the bcc(110) film, whose
surface density in the relaxed state is about 12\% higher. Strictly speaking,
the structures do not show the long range translational symmetry of a bcc
lattice. Nevertheless, it will be shown quantitatively using linear elasticity
theory that the local structure of the film is clearly bcc-derived and we will
refer to it as ``bcc-like''. Although there seems to be a large variety of
structures, all of them consist of alternating sequences of stripes of bcc-like
twins, leading to a zigzag-like deformation of the originally straight atom
rows. The different structures are distinguished mainly by the width of their
constituting bcc-like stripes. The $(1\times{}4)$ structure seen in 2.5 ML
films (2.7 ML AES) grown and imaged at 80 K consists of stripes only 2 rows
wide, which makes it most compatible with the substrate lattice---only every
second atom row is shifted by a quarter of the surface lattice constant. In
general, the ordering in this film was low and it contained a mixture of fcc
$(1\times{}1)$, $(1\times{}2)$, $(1\times{}4)$, and $(1\times{}5)$ unit cells.
The slightly more ordered structure in 3.0 ML (3.3 ML AES) films, which show
the highest bcc-like content of all studied films, significantly more than
50\%, contains \setlength{\parindent}{10pt}

\begin{figure}
\epsfxsize=0.85\hsize\epsffile[30 90 490 750]{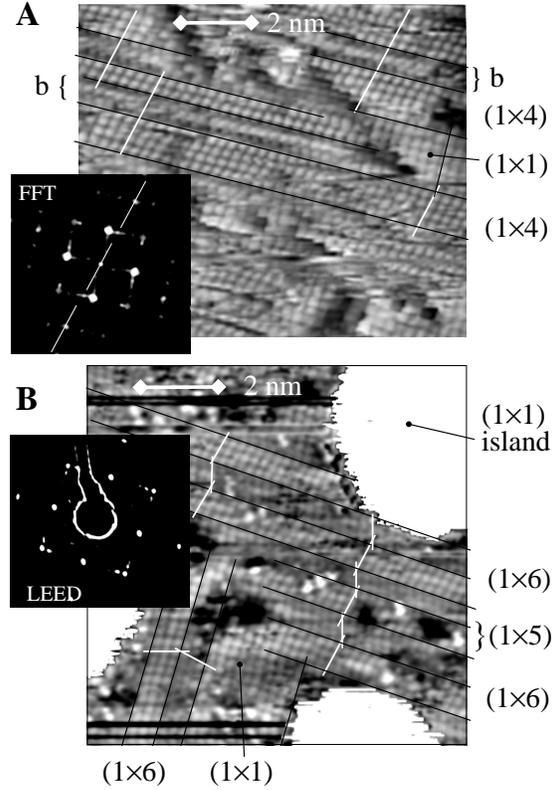} \caption{STM
images of films grown at 310 K, showing their microdomain structure. The black
lines indicate the aligned fcc$<$100$>$/bcc$<$111$>$ directions; the white
lines mark the alternate bcc$<$111$>$ directions, which are tilted by
$\sim{}14^{\circ}$ with respect to the fcc$<$100$>$ directions. (A) 3.0 ML film
(3.3 ML AES) imaged at 80 K: $(1\times{}4)$-like stripes disrupted by
boundaries ``b'' and small $(1\times{}1)$ patches. The film is predominantly
bcc-like. For details regarding the FFT image see text. (B) 4.8 ML film (4.0 ML
AES) imaged at 300 K: ($1\times{}6)$ domains embedded in a predominantly
$(1\times{}1)$ surface. The LEED pattern was acquired at 166 eV.}
\label{topology}
\end{figure}

\setlength{\parindent}{0pt} also 4 rows wide and occasionally even wider
bcc-like stripes. These wider stripes can be easily generated from the
``narrow-striped'' $(1\times{}4)$ lattice by shifting a hollow-site row by half
the surface lattice constant (arrows in Fig.~1A,B). They show indications of
local stress relaxation (see tick marks in Fig.~1B): The distance between the
atom rows appears to contract while the shear angle increases slightly to
$16^\circ$. The narrow separating boundaries (marked ``b'' in Fig.~1B,~2A)
appear blurred, which might be due to an asymmetric STM-tip or a different
crystal structure of the boundaries. In any case, these boundaries do not
disturb the in-phase relationship of neighboring equally sheared stripes
typical for the $(1\times{}4)$ structure (see filled white atoms in Fig.~1 and
white lines in Fig.~2). This is confirmed by the Fourier transformation (FFT)
of the larger STM image displayed in Fig.~2A, which shows clear $(1\times{}4)$
satellites to the principal (1,0) spots on one side. (In order to render the
FFT more like a LEED image, we artificially overlayed the FFT of the same but
$90^\circ$ rotated image to simulate a mixed rotational domain structure.) In
contrast, the $(1\times{}5)$ and $(1\times{}6)$ structures, which dominate in
thicker films, consist of bcc-like stripes two and three atom rows wide. In the
STM images of a 4.8 ML film (4.0 ML AES) displayed in Fig.~1C and 2B, the
zigzag-like bcc structure exists only in the island-free regions and comprises
only about 30\% of the total film, whereas the surface of the monoatomic
islands is pure fcc. The high fcc content indicates the close proximity to the
transition thickness between the zigzag-like bcc and the fcc phase. By
combining bcc-like stripes two and three atoms wide, which occurs in domain
boundaries of the $(1\times{}6)$ structure (see Fig.~2B), but has been seen
also in larger domains, a $(1\times{}5)$ structure is formed.
\setlength{\parindent}{10pt}

Concluding this brief structural survey, we find that while
all structures are fairly similar, the films around 3.0 ML,
which showed high magnetization in previous studies
\cite{mull95,thom92,li94}, have the highest bcc-like
content and form the widest bcc-like stripes, whereas the
films close to 2 and 5 ML show more narrow bcc-like stripes
and high fcc content, respectively, which indicates a clear
correlation of the bcc-like structure and the
magnetization. The characteristic twin structure of these
zigzag-like bcc phases combines local bcc order with a good
lattice matching at the fcc substrate interface and a
mass-transfer-free fcc-bcc transition pathway, the
hallmarks of bulk martensitic transitions (see, e.g., Ref.
\cite{waym96}), only on much smaller length scales.
Therefore, these novel bcc-like crystal structures may be
termed ``nano-martensitic bcc''.

Some aspects of the zigzag-like deformation of the fcc lattice
resemble the sinusoidal deformation model proposed by M\"uller
{\em et al.} \cite{mull95}, which was derived from I/V LEED data,
but was 2--3 times smaller in the lateral amplitude. The much
stronger zig-zag-like distortion of $\pm{}0.065$ nm (quarter of
surface lattice constant of 0.255 nm) found by us is
qualitatively different as it represents a phase transition to a
different, i.e., bcc-like crystal structure. So far, the
multi-phase microdomain structure and the incomplete
reconstruction for most film thicknesses have apparently
prevented the detection of the local bcc order by LEED
\cite{mull95}, SEXAFS \cite{magn91} or other surface averaging
experiments.

An important issue in studies of this type is the
subsurface structure of the films. STM can provide
important hints by its sensitivity for the vertical
relaxation of individual atoms (buckling). A single
mismatched layer always causes the surface atoms to assume
different vertical positions since the in-plane surface
bonds and the subsurface back-bonds are equally important.
The bcc-like stripes in our images, however, are either
very flat or show height patterns that are incompatible
with the assumption of an fcc layer directly underneath.
Typically, larger domains of the $(1\times6)$ structure
show a ``zebra-stripe" pattern (see Fig.~2B) with
alternating bright and dark bcc-like twins, each 3 atoms
wide (separated by

\begin{figure}
\epsfxsize=0.9\hsize\epsffile[30 430 570 750]{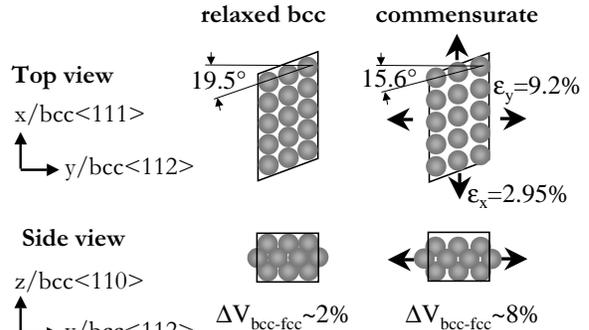} \caption {The
strained bcc film from the standpoint of linear elasticity theory: The shear
angle [with respect to the orthogonal fcc(100) lattice] is reduced to
$15.6^\circ{}$ on exposure to a lateral biaxial strain of $\varepsilon_y=9.2\%$
and $\varepsilon_x=2.95\%$, which is required to make the bcc structure
commensurate to the substrate. The volume increase of the commensurate bcc-like
film with respect to the pseudomorphic fcc film amounts to 8\%.} \label{strain}
\end{figure}

\setlength{\parindent}{0pt} black lines in Fig.~2B), with a difference in
apparent height of about 10--20 pm, which we tentatively assign to subtleties
of the stacking pattern of the zigzag-like structure.
\setlength{\parindent}{10pt}

The particular value of the shear angle of $\sim{}14^\circ$
visible in our STM images can be explained by the elastic
properties of a bcc film and emphasizes the close
relationship between the nanomartensitic bcc phase in the
2--4 ML films and the relaxed bcc structure. Figure 3
illustrates the effect of the biaxial strain of
$\varepsilon_y=9.2\%$ and $\varepsilon_x=2.95\%$, which is
required to make the bcc film commensurate, calculated by
linear elasticity theory using the anisotropic elastic
constants for bcc Fe \cite{lide97}. An immediately visible
consequence is the reduction of the shear angle from its
ideal value of $19.5^\circ$ to $15.6^\circ$, which is close
to our experimental value of $\sim{}14^\circ$. Therefore,
we conclude that the structure we observe is clearly bcc,
although significantly strained.

Previous LEED studies \cite{mull95,clar87,zhar97} also
indicate a volume expansion by about 5--6\%, which was
taken as a signature of a ferromagnetic fcc phase,
predicted for strongly strained bulk fcc Fe \cite{moru86}.
Within our structural model, however, the full volume
increase can be readily explained on the basis of linear
elasticity theory. The high in-plane strain (cf.~Fig.~3)
leads to a volume (interlayer distance) of the strained bcc
film, which is about 8\% larger than the respective value
of a commensurate fcc Fe film (1\% strained \cite{gier95}).
This agrees well with the cited LEED data, if a small fcc
admixture is included.

In order to fully appreciate the outstanding character of
the nano-martensitic bcc structures in 2--4 ML films and to
isolate possible driving forces, it is necessary to
distinguish them from the nucleation of the precursors of
the ``regular'' bcc phase in thicker films. These
precursors appear as bcc needles in fcc films more than 5
ML thick and are clearly related to the tendency of
``thick'' fcc films to assume the native bcc bulk structure
\cite{bied00nucleation}. There, the transformation to the
bcc structure occurs because the film volume is large
enough to generate sufficient energy to overcome the
barrier due to the lattice mismatch, which is created in
the course of the fcc to bcc transition. Moreover, since
these bcc needles are only 8 atom rows wide, the fcc
environment can relax the strain in the bcc needles
somewhat, shifting the energy balance in favor of the bcc
structure. In contrast, the larger defect-free bcc-like
areas in the 2-4 ML films (e.g., Fig.~2B) have only half
the volume (thickness) and additionally cannot relax the
strain equally well. To enable the phase transition in
these 2--4 ML films, an additional energy contribution is
necessary. At least two finite size effects may contribute
to this energy:

(1) Magnetic energy: It is known that surfaces and thin
films show an increased tendency towards magnetism. For
example, first principles calculations indicate a
significant increase of the magnetic moment in particular
for Fe surfaces, which contributes to their low surface
energies \cite{alde92}. Since the magnetic moment is
enhanced not only in the surface layer but also in the
layers adjacent to the Cu interface, the ``bulk'' of 2--4
ML bcc films is magnetically very different from bulk Fe
(cf. also Refs.\cite{moro99,spis00} for fcc films). Indeed,
spin-polarized x-ray appearance-potential spectroscopy
shows an increase of the spin asymmetry in 2.5 ML films on
Cu(100) of about 12\% with respect to 17 ML films or the
surface of bulk samples \cite{detz95}.

(2) Magnetic entropy: The subtle balance between fcc and
bcc in thin films is reminiscent of the entropy-driven
bcc-fcc phase transitions in bulk Fe at higher
temperatures. In bulk Fe the high temperature bcc phase
($\delta$) is stabilized above 1665 K due to the rapid
increase of magnetic disorder entropy around the bcc Curie
temperature of 1040 K \cite{zene67,hase83}. In 3 ML films
the Curie temperature is only 380 K \cite{thom92}, which
might lead to a decrease of the bcc free energy (via
disorder entropy) by a few meV \cite{hase83} in the 3 ML
room temperature films already at room temperature.

To date, the available experimental and simulation data are
insufficient to unambiguously explain the bcc-like phases
in the 2--4 ML films. Nevertheless, we favor an enhanced
magnetic energy as driving force as this contribution can
easily generate sufficient energy to compensate the
mismatch effects in the ultrathin bcc-like films.

In summary, we have observed bcc-like $(1\times{}n)$ phases
of Fe in ultrathin films below 5 ML film thickness. All
structures consist of narrow stripes of strained bcc twins
mostly 2--4 atom rows wide. This nano-martensitic bcc
structure provides a natural explanation for the previously
reported ferromagnetic character of these films. The volume
(interlayer) expansion, which originally was the key
argument to motivate a ferromagnetic fcc-like phase, is now
explained as an elastic effect due to the strongly
anisotropic strain state of the commensurate bcc-like film.
The surprising stability of a bcc-like phase in a regime
that was expected to be predominantly pseudomorphic fcc
stresses the importance of finite film thickness and/or
finite temperature effects in these ultrathin films.

We gratefully acknowledge support by the ``Fonds zur
F\"orderung der Wissenschaftlichen Forschung" (Austrian
Science Foundation) under project number Y75-PHY.

\nocite{*}
\bibliographystyle{prsty}
\bibliography{cufe100}

\end{document}